\documentclass[reviewcopy]{elsarticle}

\usepackage[reviewcopy]{adndt}
\usepackage{longtable}

\usepackage{mathptmx}

\usepackage{amsmath}

\usepackage{amssymb}

\biboptions{square,sort&compress}
\bibpunct[]{[}{]}{,}{n}{}{;}
\usepackage{xcolor}
\usepackage{multirow}
\usepackage{euscript}
\usepackage{bm}
\usepackage{mathrsfs}
\usepackage{rotating}

\newcommand{\wm}{\textcolor{white}{-}}

\begin{document}

\begin{frontmatter}

\journal{Atomic Data and Nuclear Data Tables}


\title{Spectral data for doubly excited states of helium with non-zero total angular momentum}

  \author[One,Two]{Johannes Eiglsperger\corref{cor1}}
  \ead{jeiglspe@ph.tum.de}

  \author[One]{Moritz Sch\"onwetter}

  \author[Three]{Bernard Piraux} 

  \author[One]{Javier Madro{\~n}ero}

  \cortext[cor1]{Corresponding author.}

  \address[One]{Physik Department, Technische Universit\"at M\"unchen, D-85747 Garching, Germany}
  \address[Two]{Institut f\"ur Theoretische Physik, Universit\"at Regensburg, D-93040 Regensburg, Germany}

  \address[Three]{Institute of Condensed Matter and Nanosciences, Universit\'e catholique de Louvain, B-1348 Louvain-la-Neuve, Belgium}

\date{02.08.2010} 

\begin{abstract}  
A spectral approach is used to evaluate energies and widths for a wide range of singlet and triplet resonance states of helium. Data for total angular momentum $L=1,\dots,4$ is presented for resonances up to below the 5th single ionization threshold. In addition the expectation value of $\cos(\theta_{12})$ is given for the calculated resonances.
\end{abstract}

\end{frontmatter}




\newpage

\tableofcontents
\listofDtables
\vskip5pc


\section{Introduction}

Since the discovery of strong electron-electron correlation effects in doubly excited states of helium through the seminal experiment by Madden and Codling \cite{MC:PRL10-516}, these states have attracted the continuous interest of both theoreticians and experimentalists. Direct manifestations of electronic correlations are found for instance in certain highly asymmetrically doubly excited states which are associated to highly correlated classical configurations, such as the frozen planet configuration \cite{RW:PRL65-1965,RBWS:JPB25-3929}. Studies for one-dimensional \cite{PS:Diss99,SB:EPJD22-401} and planar helium \cite{JM:Diss04,MB:PRA77-053402} suggest that these states form, under near resonant driving, non-dispersive two-electron wave packets \cite{BDZ:PRep368-409}, i.e., very robust quantum objects, which propagate along the frozen planet classical trajectory. However, the existence of these highly correlated wave packets still awaits its confirmation in full three-dimensional calculations and its experimental verification. Manifestations of electronic correlations have also been observed in double ionization of helium from the ground state by strong laser fields \cite{FBCK:PRL69-2642,WSDASK:PRL73-1227}. An enhancement by several orders of magnitude for the production of doubly charged ions is observed compared to the yield expected on basis of a single active electron approximation \cite{SYDK:PRL70-1599,YSWKAD:PRL71-3770}, in which the electron-electron interaction is neglected. This is interpreted as a fingerprint of correlated electronic ionization processes (manifesting in \emph{non-sequential} ionization, as opposed to \emph{sequential} ionization in the independent particle picture), where one electron is ``knocked out'' by the other one in a laser-induced recollision process. The geometry of the fragmentation process observed in more refined experiments \cite{W+:Nat405-658,M+:PRA65-035401} also suggests a strong dependence of the ionization process on the electronic structure \cite{dJ+:JPB37-L161} of helium-like atoms. Highly doubly excited states are expected to play an important role in the ionization by low frequency intense laser pulses \cite{PDTSBD:PRL96-133001,NGSES:PRL101-233001}. However, an accurate theoretical treatment of such a problem defines a formidable theoretical and numerical challenge due to the field induced coupling of several total angular momenta and the dimensions of the matrices associated to single total angular momenta. Note, however, that a three-dimensional ab initio fully numerical treatment of the ionization of helium in the low frequency regime is available \cite{PMMT:JPB40-1729} and has already been used to give a rather qualitative description of the correlations in the ionization process of helium from the ground state by a 780\,nm laser pulse of peak intensity $(0.275\:-\:14.4) \times 10^{14}$\,W/cm$^2$. However, due to the difficulty to extract physical information from this grid approach and its high requirements concerning computational resources, an accurate spectral approach to this problem becomes even more desirable. Further correlation effects have been observed in two-photon double ionization by strong XUV pulses where almost no experimental data is available and theoretical predictions \cite{NL:JPB34-545,FvdH:PRA66-031402R,FvdH:JPB36-L1,PBLB:EPJD26-7,HCC:JPB38-L35,IK:PRA71-043405,KI:JPB39-1731,FLKEP:PRA74-063409,F+:JPB41-051001} for the two-photon double ionization cross section among themselves deviate by orders of magnitude.

The understanding of each of these issues requires an accurate description of (highly) doubly excited states for various values of the total angular momentum $L$. The electronic correlation and the associated non-integrability of the three-body problem forces us to recur to numerical and approximation methods, which include, e.g., variational approaches, grid methods and spectral methods. Probably the most successful approaches for the description of spectral properties of two-electron atoms are spectral methods, of which two basis types can be considered: the so-called explicitly correlated bases, in which the basis functions depend explicitly on the interelectronic distance $r_{12}$, and the configuration interaction (CI) bases, in which the wave function is written as a linear combination of (antisymmetrized) products of one-electron wave functions.\footnote{See \cite{FLKEP:PRA74-063409} for a more detailed discussion.} Explicitely correlated bases allow for a very accurate description of two-electron atoms, however, the computation of the matrix elements either involves coupled three-dimensional radial integrals or is based on an analytic computation and selection rules, the number of which grows rapidly with increasing total angular momentum $L$. Moreover, rather large bases are needed for the description of highly asymmetrically excited states. Note, however, that due to the resulting analytic computation of matrix elements combined with selection rules, the explicitly correlated expansion in terms of Coulomb-Sturmian functions of the perimetric coordinates \cite{P:PRev112-1649,P:PRev115-1216} is probably the most successfull method for the treatment of highly doubly excited states with $L=0,1$ \cite{BWR:JPB28-3163,RSDK:JPB30-4663,GD:EPL40-363,BG:Diss97,JPDMK:PRA78-021401R}. Configuration interaction bases have been widely used due to their simplicity and flexibility, however, they are plagued with slow convergence for symmetrically excited states and most severely for the ground state. This is due to the fact that the basis expansion does not satisfy the Kato cusp condition associated with the coalescence of the two electrons \cite{K:CPAM10-151,W:PRev122-1826,RW:RMP32-194}. Moreover, the standard configuration interaction approach requires large basis sizes for the description of highly asymmetrically excited states. However, the computation of matrix elements in these bases involves at most two-dimensional coupled radial integrals and the computation of states with high total angular momentum $L$ does not pose any additional difficulties; these bases are frequently used for the description of few-photon ionization processes \cite{LKPS:PRA63-040502,FvdH:PRA66-031402R,FvdH:JPB36-L1,MHRM:PRA69-032707,LB:PRA69-033408,FLKEP:PRA74-063409,AFPSHNM:PRA78-023415} where highly doubly excited states do not play a fundamental role. So far configuration interaction bases have hardly been used for the computation of doubly excited states of helium, however, a non-standard member of this class has recently been shown to be capable of accurately describing autoionizing and non-autoionizing doubly excited states of helium \cite{EPM:PRA80-022511,EPM:PRA81-042527,EPM:PRA81-042528} up to the tenth single ionization threshold.

In this contribution, we present data for a large number of resonances between the first and fifth single ionization threshold with total angular momentum $L=1,\dots,4$ and parity $\pi=(-1)^L$. Results for both singlet and triplet symmetry are given. Before presenting all of our results, some of them are compared with existing data.

Unless stated otherwise atomic units (a.u.) are used throughout this contribution.

\section{Theoretical approach}

A description of our approach has already been given elsewhere \cite{GLK:Diss99,LKPS:PRA63-040502,FLKEP:PRA74-063409,EF:Diss08,EPM:PRA80-022511,EPM:PRA81-042527,EPM:PRA81-042528,JE:Diss10}. In particular the matrix representation and the numerical aspects, including efficient computation of the involved matrix elements and solution of the eigenvalue problem, have been discussed in detail in \cite{EPM:PRA80-022511,JE:Diss10}. We will thus give only a brief review of the most relevant aspects of our approach.

\subsection{Spectral method}

The nonrelativistic Hamiltonian $H$ for a two-electron atom with an infinitely heavy, pointlike nucleus of charge $Z$ is given by 
\begin{eqnarray}\label{Eq:Hamiltonian3D}
H&=&\frac{\vec{p}_{1}^{\,2}}{2}+\frac{\vec{p}_{2}^{\,2}}{2}-\frac{Z}{r_1}-\frac{Z}{r_2}+\frac{1}{r_{12}}\:,
\end{eqnarray}
with the interelectronic distance
\begin{eqnarray}\label{Eq:IntElecDis}
\frac{1}{r_{12}}&=&\frac{1}{|\vec{r}_1-\vec{r}_2|}\:,
\end{eqnarray}
and $\vec{r}_1$, $\vec{r}_2$, $\vec{p}_1$ and $\vec{p}_2$ the position and momentum vectors of particle one and two, respectively. The eigenstate wavefunction of a two-electron atom with total energy $E$ satisfies the time independent Schr\"odinger equation 
\begin{eqnarray}\label{Eq:TISE}
(H-E)\Psi(\vec{r}_{1},\vec{r}_{2})=0\:.
\end{eqnarray} 

Unlike in hydrogen, exact eigenfunctions of the Hamiltonian (\ref{Eq:Hamiltonian3D}) cannot be found. Here, in order to solve the time independent Schr\"odinger equation (\ref{Eq:TISE}) a spectral method, which consists of an expansion of the spatial wave function in terms of suitably chosen basis functions, is used. In our approach \cite{GLK:Diss99,LKPS:PRA63-040502,FLKEP:PRA74-063409,EF:Diss08,EPM:PRA80-022511,EPM:PRA81-042527,EPM:PRA81-042528,JE:Diss10} the solutions to Eq. (\ref{Eq:TISE}) are expanded as follows:
\begin{eqnarray}\label{Eq:WFExpansion}
\Psi(\vec{r}_{1},\vec{r}_{2})=\sum_{L,M}\sum_{\epsilon_{12},\pi}{\sum_{l_1,l_2}}^\pi\sum_{s}\sum_{n_1,n_2}\psi_{k_{1s},k_{2s},n_1,n_2}^{l_1,l_2,L,M,\epsilon_{12}}\beta_{n_1,n_2}^{l_1,l_2}\mathscr{A}\frac{S^{(k_{1s})}_{n_1,l_1}(r_1)}{r_1}\frac{S^{(k_{2s})}_{n_2,l_2}(r_2)}{r_2}\Lambda_{l_1,l_2}^{L,M}(\hat{r}_1,\hat{r}_2)\:,
\end{eqnarray}
where $\psi_{k_{1s},k_{2s},n_1,n_2}^{l_1,l_2,L,M,\epsilon_{12}}$ is the expansion coefficient and
\begin{eqnarray}
\beta_{n_1,n_2}^{l_1,l_2}=1+\left(\frac{1}{\sqrt{2}}-1\right)\delta_{n_1,n_2}\delta_{l_1,l_2}\:,
\end{eqnarray}
controls the redundancy that occurs within the basis due to symmetrization for basis states with equal quantum numbers for particle one and two. The symbol $\sum^\pi$ indicates that these sums depend on the parity $\pi$. The symmetry or antisymmetry of the spatial wave function, as required by the Pauli principle, is ensured by a projection onto either singlet or triplet states via the operator
\begin{eqnarray}
\mathscr{A}=\frac{1+\epsilon_{12} P_{12}}{\sqrt{2}}\:,
\end{eqnarray}
where the operator $P_{12}$ exchanges the coordinates of both electrons and $\epsilon_{12}$ takes values of $+1$ or $-1$. The radial one-electron functions $S_{n,l}^{(k)}(r)$ are Coulomb-Sturmian functions \cite{R:AAMOP6-233, HPBG:PRA55-2132} defined for a given angular momentum $l$ and radial index $n$ by
\begin{eqnarray}\label{Eq:CoulSturmFunct}
S_{n,l}^{(k)}(r)&=&N_{n,l}^{(k)}\mathrm{e}^{-kr}(2kr)^{l+1} L_{n-l-1}^{(2l+1)}(2kr)\:,
\end{eqnarray}
where $k$ is a dilation parameter, $L_{n-l-1}^{(2l+1)}(2kr)$ is an associated Laguerre polynomial and $N_{n,l}^{(k)}$ the normalization constant given by
\begin{eqnarray}
N_{n,l}^{(k)}&=&\sqrt{\frac{k}{n}}\left(\frac{(n-l-1)!}{(n+l)!}\right)^{1/2}\:.
\end{eqnarray}
The orthogonality relation for the Coulomb-Sturmian functions reads
\begin{eqnarray}\label{Eq:Orthogonality}
\int\limits_{0}^{\infty}\mathrm{d}r\:S_{n,l}^{(k)}(r)\frac{1}{r}S_{n',l}^{(k)}(r)=\frac{k}{n}\delta_{nn'}\:.
\end{eqnarray}
The radial index $n$ of the Sturmian functions is a positive integer satisfying $n\geqslant l+1$. The angular part of the expansion (\ref{Eq:WFExpansion}) is expressed in terms of bipolar spherical harmonics \cite{VMK:QuantTheoAngMom08},
\begin{eqnarray}\label{Eq:BipolarSpherHarm}
\Lambda_{l_1,l_2}^{L,M}(\hat{r}_1,\hat{r}_2)&=&\sum_{m_1,m_2}\langle l_1,m_1,l_2,m_2|L,M\rangle\: Y_{l_1,m_1}(\hat{r}_1)Y_{l_2,m_2}(\hat{r}_2)\:,
\end{eqnarray}
which couple the two individual angular momenta $l_1$ and $l_2$ in the $L-S$ scheme. $Y_{l,m}$ denotes the spherical harmonics and $\langle l_1,m_1,l_2,m_2|L,M\rangle$ is a Clebsch-Gordan coefficient. The relevant angular configurations ($l_1,l_2$) for the description of states with total angular momentum $L$ are determined by the triangle relation for the addition of angular momenta
\begin{eqnarray}
|l_1-l_2|\leq L\leq l_1+l_2\:.
\end{eqnarray}
In order to preserve parity $\pi$, which is a good quantum number, the $L-S$ coupled individual angular momenta of the electrons must satisfy $\pi=(-1)^{l_1+l_2}$. This is also reflected in $\sum^\pi$, which stands for the sum over individual angular momenta $l_1$, $l_2$ for a given parity $\pi$. The associated spin symmetry is determined by the exchange symmetry, with $\epsilon_{12}=+1$ and $\epsilon_{12}=-1$ defining singlet and triplet symmetry for $S=0$ and $S=1$, respectively. In total this allows us to target eigenstates classified by $^{2S+1}L^\pi$ and total angular momentum projection $M$ by fixing $L$, $M$, $\pi$ and $\epsilon_{12}$ in expansion (\ref{Eq:WFExpansion}).

Within a CI approach the interelectronic distance $r_{12}$ is not an explicit coordinate and therefore not accessible directly. To obtain an expression for $1/r_{12}$ in the Hamiltonian (\ref{Eq:Hamiltonian3D}) one has to exploit the multipole expansion of the electron-electron repulsion:
\begin{eqnarray}\label{Eq:IntElDisMultExp}
\frac{1}{r_{12}}&=&\sum_{q=0}^{\infty}\sum_{p=-q}^{q}\frac{4\pi}{2q+1}\frac{r_{<}^{q}}{r_{>}^{q+1}}Y_{q,p}^{*}(\hat{r}_1)Y_{q,p}(\hat{r}_2)\:,
\end{eqnarray}
with $r_{<}=\textrm{min}(r_1,r_2)$ and $r_{>}=\textrm{max}(r_1,r_2)$.

In general the CI expansions involving Coulomb-Sturmian functions use the same dilation parameter $k$ for all Coulomb-Sturmian functions, which is equivalent to setting $k_{1s}=k_{2s}\equiv k$ and $s=1$ in our expansion (\ref{Eq:WFExpansion}). Furthermore, for each pair of $(l_1,l_2)$, the same number $N$ of Coulomb-Sturmian functions $S_{n_1,l_1}^{(k)}(r_1)$ with $l_1+1\leqslant n_1 \leqslant l_1+N$ and $S_{n_2,l_2}^{(k)}(r_2)$ with $l_2+1\leqslant n_2 \leqslant l_2+N$ is chosen for the representation.
In contrast, our approach is constructed in order to allow the dilation parameter and the number of Coulomb-Sturmian functions associated to one electron to be different from those attributed to the other electron. This leads to the introduction of a set of Coulomb-Sturmian functions $\{S_{n_1,l_1}^{(k_{1s})}(r_1),\,S_{n_2,l_2}^{(k_{2s})}(r_2)\}$ associated to electron one and two, which is characterized by the combination $[k_{1s},N_{1s}^\textrm{min},N_{1s}^\textrm{max},k_{2s},N_{2s}^\textrm{min},N_{2s}^\textrm{max}]$ 
with \hbox{$l_1+N_{1s}^\textrm{min}\leqslant n_1 \leqslant l_1+N_{1s}^\textrm{max}$} and $l_2+N_{2s}^\textrm{min}\leqslant n_2 \leqslant l_2+N_{2s}^\textrm{max}$. Moreover, more than one and different sets -- labeled by the subscript $s$ -- may be selected for any angular configuration $(l_1,l_2)$. To avoid redundancies in expansion (\ref{Eq:WFExpansion}), the orbital angular momenta are restricted to $l_1\leqslant l_2$, and if $l_1=l_2$ and $k_{1s}=k_{2s}$ to $n_1\leqslant n_2$. Be aware of the fact, that due to the restriction to $l_1\leqslant\index{} l_2$, each set of Coulomb-Sturmian functions $[k_{1,s},N_{1,s}^\textrm{min},N_{1,s}^\textrm{max},k_{2,s},N_{2,s}^\textrm{min},N_{2,s}^\textrm{max}]$ should be accompanied by $[k_{2,s},N_{2,s}^\textrm{min},N_{2,s}^\textrm{max},k_{1,s},N_{1,s}^\textrm{min},N_{1,s}^\textrm{max}]$ in the case of $k_{1,s}\neq k_{2,s}$ and $l_1\neq l_2$. The reason for this is that, e.g., sets with $k_{1,s}>k_{2,s}$ would explicitly favour a smaller extent of the $l_1$-orbital than of the $l_2$-orbital. This would limit the descriptive power of the basis after truncation. To illustrate the importance of this kind of symmetrization let us consider states of $L=1$ below the second single ionization threshold. In the independent particle model the spectrum consists of $2snp$, $2pns$ and $2pnd$ states. Using only sets with $k_{1,s}>k_{2,s}$ would allow a good representation of $2snp$ and $2pnd$ states, however, the description of the $2pns$ would be very poor in a truncated basis. As already realized by the first experiment on doubly excited states \cite{MC:PRL10-516} and its theoretical interpretation \cite{CFP:PRL10-518} the electron-electron interaction mixes the different configurations of the independent particle model. Consequently, the exclusive use of sets with $k_{1,s}>k_{2,s}$ requires a huge basis to get converged results, which is not the case if the mirrored set of Coulomb-Sturmian functions $[k_{2,s},N_{2,s}^\textrm{min},N_{2,s}^\textrm{max},k_{1,s},N_{1,s}^\textrm{min},N_{1,s}^\textrm{max}]$ is included.

Combined with the complex rotation method presented in section \ref{Sec:ComplexRotation} the approach allows the description of resonance states. By choosing appropriate sets of Coulomb-Sturmian functions the description of a given energy regime, i.e., below a certain ionization threshold, is possible with a rather small number of basis functions \cite{EPM:PRA80-022511}. This is in particular true for highly asymmetrically excited states.

\subsection{Complex rotation}\label{Sec:ComplexRotation}

The electron-electron interaction in helium couples different channels of the non-interacting two-electron dynamics, and gives rise to resonance states embedded in the continua above the first single ionization threshold. To extract the energies and decay rates of resonance states we use complex rotation (or ``dilation'') \cite{AC:CMP22-269, BC:CMP22-280, S:AoM97-247, R:ARPC33-223, H:PRep99-1}, which was shown to be applicable for the Coulomb potential in \cite{GGS:AIHP42-215}.

The complex dilation of any operator by an angle $\theta$ is mediated by the non-unitary complex rotation operator 
\begin{equation}\label{Eq:CompRotOp}
 R(\theta)=\exp\left(-\theta\frac{\vec{r}\cdot\vec{p}+\vec{p}\cdot\vec{r}}2\right)\:,
\end{equation}
where $\vec{r}$ and $\vec{p}$ represent the $2N$ component vector made up of $\vec{r}_1$, $\vec{r}_2$ and $\vec{p}_1$, $\vec{p}_2$, respectively, with $N$ the dimension of the treated system. The transformation of the position and momentum operators consists of a rotation by $\theta$ in the complex plane,
\begin{eqnarray}
\vec{r}& \rightarrow &R(\theta)\,\vec{r}\, R(-\theta)=\vec{r}\, \mathrm{e}^{\mathrm{i}\theta}\:,\nonumber\\
\vec{p}& \rightarrow &R(\theta)\,\vec{p}\, R(-\theta)=\vec{p}\, \mathrm{e}^{-\mathrm{i}\theta}\:.
\end{eqnarray}
Thus, the Hamiltonian (\ref{Eq:Hamiltonian3D}) transforms into,
\begin{eqnarray}\label{Eq:CRotHamiltonian}
H(\theta)=R(\theta)HR(-\theta)=\left(\frac{\vec{p}_{1}^{\,2}+\vec{p}_{2}^{\,2}}{2}\right)\mathrm{e}^{-2\mathrm{i}\theta}-\left(\frac{Z}{r_1}+\frac{Z}{r_2}-\frac{1}{r_{12}}\right)\mathrm{e}^{-\mathrm{i}\theta}.
\end{eqnarray}
This operator is no longer Hermitean and, therefore, its eigenvalues are in general complex. However, the spectrum of the rotated Hamiltonian is related to the spectrum of the unrotated operator according to \cite{BC:CMP22-280, R:ARPC33-223, GGS:AIHP42-215}:
\begin{enumerate}
\item The bound spectrum of $H$ is invariant under the complex rotation.
\item The continuum states are located on half lines, rotated by an angle $-2\theta$ around the ionization thresholds of the unrotated Hamiltonian into the lower half of the complex plane.
\item There are isolated complex eigenvalues \hbox{$E_{i,\theta}=E_i-{\rm i}\Gamma_i/2$} in the lower half plane, corresponding to resonance states. These are stationary under changes of $\theta$, provided the dilation angle is large enough to uncover their positions on the Riemannian sheets of the associated resolvent \cite{PS:PRA43-3764,SR:AnalysisOfOperators78}. The associated resonance eigenfunctions are square integrable \cite{H:PRep99-1}, in contrast to the resonance eigenfunctions of the unrotated Hamiltonian.
\end{enumerate}

The eigenstates of $H(\theta)$,
\begin{eqnarray}
H(\theta)|\Psi_{i,\theta}\rangle = E_{i,\theta}|\Psi_{i,\theta}\rangle \:, 
\end{eqnarray}
are normalized for the scalar product
\begin{eqnarray}\label{Eq:NormCmplRot}
\langle\Psi_{j,-\theta}|\Psi_{i,\theta}\rangle = \delta_{ij} \:,
\end{eqnarray}
and satisfy the closure relation:
\begin{eqnarray}
\sum_i |\Psi_{i,\theta}\rangle\langle\Psi_{i,-\theta}|=1\:.
\end{eqnarray} 
Following \cite{BGD:JPB27-2663}, the Green function of the rotated Hamiltonian reads:
\begin{eqnarray}\label{Eq:GreenFuncRotHam}
G_\theta=\frac{1}{E-H(\theta)}=\sum_i \frac{|\Psi_{i,\theta}\rangle\langle\Psi_{i,-\theta}|}{E-E_{i,\theta}}\:,
\end{eqnarray}
while the relation between the Green function of the unrotated Hamiltonian and Eq. (\ref{Eq:GreenFuncRotHam}) has been shown \cite{JR:PRA28-1930} to be:
\begin{eqnarray}\label{Eq:RelationGreenFuncUnrotRot}
G(E)&=&\frac{1}{E-H}=R(-\theta)G_{\theta}(E)R(\theta)\:.
\end{eqnarray}
The projection operator on a real energy eigenstate is related to the Green function through
\begin{eqnarray}\label{Eq:ProjectOp}
|\phi_E\rangle\langle\phi_E|=\frac{1}{2\mathrm{i}\pi}\left(G^{-}(E)-G^{+}(E)\right)\:,
\end{eqnarray}
with
\begin{eqnarray}
G^{\pm}(E)=\frac{1}{E\pm\mathrm{i}\eta-H}\:,\quad \eta\rightarrow0^+\:.
\end{eqnarray}
Using Eq. (\ref{Eq:GreenFuncRotHam}) and Eq. (\ref{Eq:RelationGreenFuncUnrotRot}) gives for the projection operator on a real energy eigenstate, in terms of the eigenstates of the rotated Hamiltonian,
\begin{eqnarray}\label{Eq:ProjectOpComplRot}
|\phi_E\rangle\langle\phi_E|\!\!\!&=&\!\!\!\frac{1}{2\mathrm{i}\pi}
\sum_i\Bigg[\frac{R(-\theta)|\Psi_{i,\theta}\rangle\langle\Psi_{i,-\theta}|R(\theta)}{E_{i,\theta}-E}-\frac{R(\theta)|\Psi_{i,-\theta}\rangle\langle\Psi_{i,\theta}|R(-\theta)}{E_{i,-\theta}-E}\Bigg]\:.
\end{eqnarray}

\subsection{Expectation value of $\cos(\theta_{12})$}

The expectation value of $\cos(\theta_{12})$, where $\theta_{12}$ is the mutual angle between the
position vectors of the electrons, for a given state $\phi_E$ of energy E is obtained, up to normalization of $|\phi_E\rangle$, by
\begin{eqnarray}
\left\langle\phi_E|\cos(\theta_{12})|\phi_E\right\rangle&=&\frac{1}{2\pi\mathrm{i}}\sum_i\left[\frac{\langle\Psi_{i,-\theta}|\cos(\theta_{12})|\Psi_{i,\theta}\rangle}{E_{i,\theta}-E}-\frac{\langle\Psi_{i,\theta}|\cos(\theta_{12})|\Psi_{i,-\theta}\rangle}{E_{i,-\theta}-E}\right]\nonumber\\
&=&\frac{1}{\pi}\textrm{Im}\left[\sum_i\frac{\langle\overline{\Psi_{i,\theta}}|\cos(\theta_{12})|\Psi_{i,\theta}\rangle}{E_{i,\theta}-E}\right]\:,
\end{eqnarray}
where we have used the projector (\ref{Eq:ProjectOpComplRot}). Similarly, the square of the norm of $|\phi_E\rangle$ reads as
\begin{eqnarray}
\left\langle\phi_E|\phi_E\right\rangle&=&\frac{1}{2\pi\mathrm{i}}\sum_i\left[\frac{\langle\Psi_{i,-\theta}|\Psi_{i,\theta}\rangle}{E_{i,\theta}-E}-\frac{\langle\Psi_{i,\theta}|\Psi_{i,-\theta}\rangle}{E_{i,-\theta}-E}\right]\nonumber\\
&=&\frac{1}{\pi}\textrm{Im}\left[\sum_i\frac{1}{E_{i,\theta}-E}\right]\:,
\end{eqnarray}
where the normalization (\ref{Eq:NormCmplRot}) of $|\Psi_{i,\theta}\rangle$ has been taken into account in the last step. The expectation value is thus given by
\begin{eqnarray}
\langle\cos(\theta_{12})\rangle=\frac{\left\langle\phi_E|\cos(\theta_{12})|\phi_E\right\rangle}{\left\langle\phi_E|\phi_E\right\rangle}=\frac{\displaystyle\textrm{Im}\left[\displaystyle\sum\limits_i\frac{\displaystyle\langle\overline{\Psi_{i,\theta}}|\cos(\theta_{12})|\Psi_{i,\theta}\rangle}{\displaystyle E_{i,\theta}-E}\right]}{\displaystyle\textrm{Im}\left[\displaystyle\sum\limits_i\frac{\displaystyle1}{\displaystyle E_{i,\theta}-E}\right]}\:.
\end{eqnarray}
A well isolated resonance  $|\Psi_{j,\theta}\rangle$ with $E_{j,\theta}\simeq E$ and $\left|\mathrm{Re}(E_{j,\theta})-\mathrm{Re}(E_{i,\theta})\right|\gg\left|E_{j,\theta}-E\right|$, $\forall i\neq j$, gives the dominant contribution to the above sum, and justifies the {\em single pole approximation} \cite{BGD:JPB27-2663}
\begin{eqnarray}
\left\langle\phi_E|\cos(\theta_{12})|\phi_E\right\rangle&\simeq&\frac{1}{\pi|\textrm{Im}(E_{i,\theta})|}\textrm{Re}\langle\overline{\Psi_{i,\theta}}|\cos(\theta_{12})|\Psi_{i,\theta}\rangle\:,\\
\left\langle\phi_E|\phi_E\right\rangle&\simeq&\frac{1}{\pi|\textrm{Im}(E_{i,\theta})|}\:,\label{Eq:SinglePoleNormCmplxRot}
\end{eqnarray}
leading to
\begin{eqnarray}
\langle\cos(\theta_{12})\rangle\simeq\textrm{Re}(\langle\overline{\Psi_{i,\theta}}|\cos(\theta_{12})|\Psi_{i,\theta}\rangle)\:.
\end{eqnarray}
Rewriting $\cos(\theta_{12})$ in terms of spherical harmonics $Y_{l,m}(\hat{r})$,
\begin{eqnarray}\label{Eq:costheta12-3d}
\cos(\theta_{12})=\frac{\vec{r}_1\cdot \vec{r}_2}{r_1r_2}=\frac{4\pi}{3}\sum_{q_1}\sum_{q_2} \mathcal{C}_{1,q_1,1,q_2}^{1,0} Y_{1,q_1}(\hat{r}_1) \:Y_{1,q_2}(\hat{r}_2)\:,
\end{eqnarray}
with Clebsch-Gordan coefficients $\mathcal{C}_{1,q_1,1,q_2}^{1,0}$, leads to an easily accessible matrix representation of $\cos(\theta_{12})$ -- the details of which are given in \cite{JE:Diss10} -- in terms of overlap integrals of Coulomb-Sturmian functions and Wigner $3jm$ and $6j$ symbols \cite{VMK:QuantTheoAngMom08}.

\section{Results}
Our approach was used for the computation of $^{1,3}S^\text{e}$ and $^{1,3}P^\text{e}$ resonances  in \cite{EPM:PRA80-022511,EPM:PRA81-042528} and for non-autoionizing states with various values of the angular momentum $L$ in \cite{LKPS:PRA63-040502,FLKEP:PRA74-063409,EPM:PRA81-042527}. Here we present energy, half-width and expectation value of $\cos(\theta_{12})$ for doubly excited states of helium for total angular momentum $L=1,\dots,4$ and parity $\pi=(-1)^L$. Data for both singlet and triplet symmetry are given for states from below the second up to below the fifth single ionization threshold. All results have been tested for convergence with respect to variation of basis size (including variation of the number of Sturmians as well as the number of angular configurations), of the dilation parameters and of the complex rotation angle. Only converged digits are given in the tables.

Various authors, using a wide range of different methods, have published results for these states (see e.g., \cite{BT:PRA11-2018,H:JPB12-387,H:PLA79-44,H:PRA23-2137,H:JPB15-L691,BBFT:JPB15-L603,BT:PRA29-1895,HC:JPB18-3481,O:PRA33-824,H:ZfPD11-277,SCK:PRA39-5111,HH:JPB22-3397,GB:JPB23-365,FI:JPB23-679,WX:JPB23-727,SS:PRA42-2562,H:ZfPD21-191,BMRY:ADNDT48-167,CT:PRA44-232,MS:PRA44-7206,TWM:PRA46-2437,C:PRA47-705,L:PRA49-4473,TS:PRA50-1321,DSRKW:PRA53-1424,H:ZfPD42-77,C:PRA56-4537,RSDK:JPB30-4663,AM:JPB40-3655} for  $^1P^\textrm{o}$ states, \cite{BT:PRA11-2018,H:JPB12-387,H:PLA79-44,H:JPB15-L691,H:ZfPD11-277,H:PRA23-2137,H:JPB17-L695,HC:JPB18-3481,O:PRA33-824,WX:JPB23-727,BMRY:ADNDT48-167,SS:PRA42-2562,MS:PRA44-7206,TWM:PRA46-2437,C:PRA47-705,SM:PRA47-1878,H:PRA48-3598,L:PRA49-4473,H:ZfPD42-77,C:PRA56-4537,AM:JPB40-3655,A:ADNDT94-903} for $^3P^\textrm{o}$ states, \cite{BT:PRA11-2018,HC:JPB18-3481,O:PRA33-824,H:ZfPD11-277,FI:JPB23-679,BMRY:ADNDT48-167,HB:PRA44-2895,TWM:PRA46-2437,L:PRA49-4473,C:PRA56-4537,AM:JPB40-3655} for $^1D^\textrm{e}$ states, \cite{BT:PRA11-2018,HC:JPB18-3481,O:PRA33-824,H:ZfPD11-277,BMRY:ADNDT48-167,HB:PRA44-2895,TWM:PRA46-2437,L:PRA49-4473,C:PRA56-4537,AM:JPB40-3655,A:ADNDT94-903} for $^3D^\textrm{e}$ states, \cite{HC:JPB18-3481,H:ZfPD11-277,BMRY:ADNDT48-167,L:PRA49-4473} for $^1F^\textrm{o}$ states, \cite{HC:JPB18-3481,H:ZfPD11-277,BMRY:ADNDT48-167,L:PRA49-4473} for $^3F^\textrm{o}$ states, \cite{HC:JPB18-3481,H:ZfPD11-277,BMRY:ADNDT48-167,L:PRA49-4473} for $^1G^\textrm{e}$ states and \cite{HC:JPB18-3481,H:ZfPD11-277,BMRY:ADNDT48-167} for $^3G^\textrm{e}$ states), however, most of these publications contain only a small number of resonances and results are few in general for $L\geq2$. In tables \ref{TabA} to \ref{TabH} a comparison of our results to data of some of the available data is presented, with up to 30 resonances per symmetry and threshold. For $^3P^\textrm{o}$, $^1D^\textrm{e}$, $^1F^\textrm{o}$, $^3F^\textrm{o}$, $^3G^\textrm{e}$ there are in total seven resonances below the fifth SIT available in the literature for which we do not give a result. These resonances are lying slightly above the fourth SIT and have a relatively large width. In our complex rotation calculation these are for some complex rotation angles still masked by or are still strongly influenced by the rotated continuum and we do not dare to give a result. Note, that in Ref. \cite{A:ADNDT94-903} the number of given digits differs from the actual precision. Due to conversion to atomic units and to half-widths the number of presented digits might differ slightly from those given in the respective references. For energies above the third single ionization threshold overlapping of series converging to distinct ionization thresholds is possible as for instance shown for the spectrum starting from below $I_4$ for $^3P\textrm{o}$, $^3D\textrm{e}$ \cite{AM:JPB40-3655} and from below $I_5$ for $^1P\textrm{o}$ \cite{RSDK:JPB30-4663}. A perturber is a state that belongs from its quantum numbers to a Rydberg series converging to a single ionization threshold $I_N$, however, its energy is so low that it lies below the energy of a lower threshold $I_{N-1}$. The effect of these perturbers is present in our calculation, however, given the focus of this contribution, no effort for the classification of these states has been made. Finally, in tables \ref{Tab1} to \ref{Tab8} our complete results are given. In addition, in order to provide an additional large scale comparison, our results for $^1P^\textrm{o}$ are displayed together with the highly accurate data of \cite{RSDK:JPB30-4663} in table \ref{Tab1}.

{\centering
{
\footnotesize
\begin{longtable}{ll|ll||ll|ll}
\caption{\label{TabA} $^1P^\textrm{o}$ resonances of helium from below the $N=2$ to below the $N=5$ threshold compared with other theoretical results.}\vspace{-23pt}\\
\hline
\multicolumn{2}{c|}{This work}&\multicolumn{2}{c||}{other results}&\multicolumn{2}{c|}{This work (continued)}&\multicolumn{2}{c}{other results (continued)}\\
$\textrm{Re}(-E)$&$\textrm{Im}(-E)$&$\textrm{Re}(-E)$&$\textrm{Im}(-E)$&$\textrm{Re}(-E)$&$\textrm{Im}(-E)$&$\textrm{Re}(-E)$&$\textrm{Im}(-E)$\\
\hline
\endfirsthead\\
\caption[]{(continued)}\vspace{-24pt}\\
\multicolumn{2}{c|}{This work}&\multicolumn{2}{c||}{other results}&\multicolumn{2}{c|}{This work (continued)}&\multicolumn{2}{c}{other results (continued)}\\
$\textrm{Re}(-E)$&$\textrm{Im}(-E)$&$\textrm{Re}(-E)$&$\textrm{Im}(-E)$&$\textrm{Re}(-E)$&$\textrm{Im}(-E)$&$\textrm{Re}(-E)$&$\textrm{Im}(-E)$\\
\hline
\endhead
\multicolumn{8}{c}{$N=2$}\\
\hline
0.69313     & 0.00069            & 0.693069$^\textrm{a)}$   & 0.000686$^\textrm{a)}$          &0.54649297  & 0.00000096         & 0.546490$^\textrm{a)}$   & 0.00000101$^\textrm{a)}$        \\
            &                    & 0.693135$^\textrm{b)}$   & 0.0006865$^\textrm{b)}$         &            &                    & 0.54642$^\textrm{c)}$    & 0.0000010$^\textrm{c)}$         \\
            &                    & 0.69305$^\textrm{c)}$    & 0.000687$^\textrm{c)}$          &0.5343622   & 0.0000639          & 0.534358$^\textrm{a)}$   & 0.000064$^\textrm{a)}$          \\
0.597073801 & 0.000001922        & 0.597074$^\textrm{a)}$   & 0.00000192$^\textrm{a)}$        &            &                    & 0.53429$^\textrm{c)}$    & 0.000064$^\textrm{c)}$          \\
            &                    & 0.59707$^\textrm{b)}$    & 0.00000192$^\textrm{b)}$        &0.52761598  & $<0.7\times10^{-7}$& 0.527613$^\textrm{a)}$   & 0.0000000015$^\textrm{a)}$      \\
            &                    & 0.59700$^\textrm{c)}$    & 0.00000193$^\textrm{c)}$        &            &                    & 0.52753$^\textrm{c)}$    & 0.00000000007$^\textrm{c)}$     \\
0.5640830   & 0.0001501          & 0.564074$^\textrm{a)}$   & 0.0001499$^\textrm{a)}$         &0.52729761  & 0.00000046         & 0.527295$^\textrm{a)}$   & 0.000000494$^\textrm{a)}$       \\
\end{longtable}}}
{\centering
{\footnotesize
\begin{longtable}{ll|ll||ll|ll}
\caption{\label{TabB} $^3P^\textrm{o}$ resonances of helium from below the $N=2$ to below the $N=5$ threshold compared with other theoretical results.}\vspace{-23pt}\\
\hline
\multicolumn{2}{c|}{This work}&\multicolumn{2}{c||}{other results}&\multicolumn{2}{c|}{This work (continued)}&\multicolumn{2}{c}{other results (continued)}\\
$\textrm{Re}(-E)$&$\textrm{Im}(-E)$&$\textrm{Re}(-E)$&$\textrm{Im}(-E)$&$\textrm{Re}(-E)$&$\textrm{Im}(-E)$&$\textrm{Re}(-E)$&$\textrm{Im}(-E)$\\
\hline
\endfirsthead\\
\caption[]{(continued)}\vspace{-24pt}\\
\multicolumn{2}{c|}{This work}&\multicolumn{2}{c||}{other results}&\multicolumn{2}{c|}{This work (continued)}&\multicolumn{2}{c}{other results (continued)}\\
$\textrm{Re}(-E)$&$\textrm{Im}(-E)$&$\textrm{Re}(-E)$&$\textrm{Im}(-E)$&$\textrm{Re}(-E)$&$\textrm{Im}(-E)$&$\textrm{Re}(-E)$&$\textrm{Im}(-E)$\\
\hline
\endhead
\multicolumn{8}{c}{$N=2$}\\
\hline
0.7604922   & 0.0001492           & 0.760490639$^{\textrm{a)}}$   & 0.00014945$^{\textrm{a)}}$      & 0.51710730  & 0.00000406          & 0.517107078$^{\textrm{a)}}$   & 0.000004067$^{\textrm{a)}}$     \\
            &                     & 0.760489$^{\textrm{b)}}$      & 0.0001495$^{\textrm{b)}}$       &             &                     & 0.517107$^{\textrm{b)}}$      & 0.0000043$^{\textrm{b)}}$       \\
0.5846722   & 0.0000411           & 0.584671760$^{\textrm{a)}}$   & 0.00004115$^{\textrm{a)}}$      & 0.516079739 & 0.000000118         & 0.516079530$^{\textrm{a)}}$   & 0.00000011925$^{\textrm{a)}}$   \\
            &                     & 0.584671$^{\textrm{b)}}$      & 0.0000412$^{\textrm{b)}}$       &             &                     & 0.516079$^{\textrm{b)}}$      & 0.000000115$^{\textrm{b)}}$     \\
0.57903096  & 0.00000094          & 0.579030867$^{\textrm{a)}}$   & 0.0000009475$^{\textrm{a)}}$    & 0.513156015 & $<0.4\times10^{-8}$ & 0.513155641$^{\textrm{a)}}$   & 0.0000000008725$^{\textrm{a)}}$ \\
            &                     & 0.579030$^{\textrm{b)}}$      & 0.000000925$^{\textrm{b)}}$     &             &                     & 0.513155$^{\textrm{b)}}$      & 0.0000000008$^{\textrm{b)}}$    \\
0.54884429  & $<0.1\times10^{-7}$ & 0.548844002$^{\textrm{a)}}$   & 0.000000006265$^{\textrm{a)}}$  & 0.512191320 & 0.000002448         & 0.512191135$^{\textrm{a)}}$   & 0.000002452$^{\textrm{a)}}$     \\
\end{longtable}}}
{\centering
{
\footnotesize
\begin{longtable}{ll|ll||ll|ll}
\caption{\label{TabC} $^1D^\textrm{e}$ resonances of helium from below the $N=2$ to below the $N=5$ threshold compared with other theoretical results.}\vspace{-23pt}\\
\hline
\multicolumn{2}{c|}{This work}&\multicolumn{2}{c||}{other results}&\multicolumn{2}{c|}{This work (continued)}&\multicolumn{2}{c}{other results (continued)}\\
$\textrm{Re}(-E)$&$\textrm{Im}(-E)$&$\textrm{Re}(-E)$&$\textrm{Im}(-E)$&$\textrm{Re}(-E)$&$\textrm{Im}(-E)$&$\textrm{Re}(-E)$&$\textrm{Im}(-E)$\\
\hline
\endfirsthead\\
\caption[]{(continued)}\vspace{-24pt}\\
\multicolumn{2}{c|}{This work}&\multicolumn{2}{c||}{other results}&\multicolumn{2}{c|}{This work (continued)}&\multicolumn{2}{c}{other results (continued)}\\
$\textrm{Re}(-E)$&$\textrm{Im}(-E)$&$\textrm{Re}(-E)$&$\textrm{Im}(-E)$&$\textrm{Re}(-E)$&$\textrm{Im}(-E)$&$\textrm{Re}(-E)$&$\textrm{Im}(-E)$\\
\hline
\endhead
\multicolumn{8}{c}{$N=2$}\\
\hline
0.70193     & 0.00118             & 0.70183$^{\textrm{a)}}$     & 0.00118$^{\textrm{a)}}$      & 0.52929296  & $<0.1\times10^{-7}$ & 0.529292$^{\textrm{a)}}$    & 0.00000000605$^{\textrm{a)}}$\\
            &                     & 0.7019457$^{\textrm{b)}}$   & 0.0011811$^{\textrm{b)}}$    &             &                     & 0.529292995$^{\textrm{b)}}$ & 0.0000000065$^{\textrm{b)}}$ \\
            &                     & 0.701655$^{\textrm{c)}}$    & 0.001205$^{\textrm{c)}}$     &             &                     & 0.52928885$^{\textrm{c)}}$  & 0.00000000625$^{\textrm{c)}}$\\
0.569216    & 0.000277            & 0.569193$^{\textrm{a)}}$    & 0.000280$^{\textrm{a)}}$     & 0.5227417   & 0.0000580           & 0.522737$^{\textrm{a)}}$    & 0.000059$^{\textrm{a)}}$     \\
            &                     & 0.5692205$^{\textrm{b)}}$   & 0.0002775$^{\textrm{b)}}$    &             &                     & 0.52272$^{\textrm{c)}}$     & 0.0000595$^{\textrm{c)}}$    \\
            &                     & 0.569115$^{\textrm{c)}}$    & 0.000285$^{\textrm{c)}}$     & 0.52011736  & 0.00000310          & 0.520114$^{\textrm{a)}}$    & 0.00000320$^{\textrm{a)}}$   \\
0.5564290   & 0.0000097           & 0.556417$^{\textrm{a)}}$    & 0.00001005$^{\textrm{a)}}$   &             &                     & 0.5201105$^{\textrm{c)}}$   & 0.00000315$^{\textrm{c)}}$   \\
\end{longtable}}} 

{\centering
{
\footnotesize
\begin{longtable}{ll|ll||ll|ll}
\caption{\label{TabD} $^3D^\textrm{e}$ resonances of helium from below the $N=2$ to below the $N=5$ threshold compared with other theoretical results.}\vspace{-23pt}\\
\hline
\multicolumn{2}{c|}{This work}&\multicolumn{2}{c||}{other results}&\multicolumn{2}{c|}{This work (continued)}&\multicolumn{2}{c}{other results (continued)}\\
$\textrm{Re}(-E)$&$\textrm{Im}(-E)$&$\textrm{Re}(-E)$&$\textrm{Im}(-E)$&$\textrm{Re}(-E)$&$\textrm{Im}(-E)$&$\textrm{Re}(-E)$&$\textrm{Im}(-E)$\\
\hline
\endfirsthead\\
\caption[]{(continued)}\vspace{-24pt}\\
\multicolumn{2}{c|}{This work}&\multicolumn{2}{c||}{other results}&\multicolumn{2}{c|}{This work (continued)}&\multicolumn{2}{c}{other results (continued)}\\
$\textrm{Re}(-E)$&$\textrm{Im}(-E)$&$\textrm{Re}(-E)$&$\textrm{Im}(-E)$&$\textrm{Re}(-E)$&$\textrm{Im}(-E)$&$\textrm{Re}(-E)$&$\textrm{Im}(-E)$\\
\hline
\endhead
\multicolumn{8}{c}{$N=2$}\\
\hline
0.58378424   & $<0.6\times10^{-7}$  & 0.58378427$^{\textrm{b)}}$  & 0.0000000143$^{\textrm{b)}}$       &0.509847973  & $<0.3\times10^{-8}$  & 0.509847831$^{\textrm{a)}}$ & 0.00000000000000600$^{\textrm{a)}}$\\
             &                      & 0.583784086$^{\textrm{a)}}$ & 0.00000001419$^{\textrm{a)}}$      &0.5089450531 & $<0.5\times 10^{-9}$ & 0.508944928$^{\textrm{a)}}$ & 0.0000000003949$^{\textrm{a)}}$    \\
0.56068624   & 0.00000371           & 0.5606865$^{\textrm{b)}}$   & 0.00000375$^{\textrm{b)}}$         &0.508085781  & 0.000000259          & 0.508085701$^{\textrm{a)}}$ & 0.0000002609$^{\textrm{a)}}$       \\
             &                      & 0.560686118$^{\textrm{a)}}$ & 0.000003760$^{\textrm{a)}}$        &0.507574151  & $<0.2\times10^{-8}$  & 0.507574056$^{\textrm{a)}}$ & 0.0000000000001935$^{\textrm{a)}}$ \\
0.541678994  & $<0.4\times10^{-8}$  & 0.541678920$^{\textrm{a)}}$ & 0.000000004889$^{\textrm{a)}}$     &0.5069578476 & $<0.4\times 10^{-9}$ & 0.506957765$^{\textrm{a)}}$ & 0.0000000002658$^{\textrm{a)}}$    \\
0.53346309   & 0.00000189           & 0.533463037$^{\textrm{a)}}$ & 0.0000019105$^{\textrm{a)}}$       &0.506364305  & 0.000000182          & 0.506364245$^{\textrm{a)}}$ & 0.0000001837$^{\textrm{a)}}$       \\
0.529312233  & $<0.6\times10^{-8}$  & 0.529312204$^{\textrm{a)}}$ & 0.00000000003531$^{\textrm{a)}}$   &0.506005498  & $<0.2\times10^{-8}$  & 0.506005427$^{\textrm{a)}}$ & 0.0000000000003590$^{\textrm{a)}}$ \\
\end{longtable}}}  

{\centering
{
\footnotesize
\begin{longtable}{ll|ll||ll|ll}
\caption{\label{TabE} $^1F^\textrm{o}$ resonances of helium from below the $N=3$ to below the $N=5$ threshold compared with other theoretical results.}\vspace{-23pt}\\
\hline
\multicolumn{2}{c|}{This work}&\multicolumn{2}{c||}{other results}&\multicolumn{2}{c|}{This work (continued)}&\multicolumn{2}{c}{other results (continued)}\\
$\textrm{Re}(-E)$&$\textrm{Im}(-E)$&$\textrm{Re}(-E)$&$\textrm{Im}(-E)$&$\textrm{Re}(-E)$&$\textrm{Im}(-E)$&$\textrm{Re}(-E)$&$\textrm{Im}(-E)$\\
\hline
\endfirsthead\\
\caption[]{(continued)}\vspace{-24pt}\\
\multicolumn{2}{c|}{This work}&\multicolumn{2}{c||}{other results}&\multicolumn{2}{c|}{This work (continued)}&\multicolumn{2}{c}{other results (continued)}\\
$\textrm{Re}(-E)$&$\textrm{Im}(-E)$&$\textrm{Re}(-E)$&$\textrm{Im}(-E)$&$\textrm{Re}(-E)$&$\textrm{Im}(-E)$&$\textrm{Re}(-E)$&$\textrm{Im}(-E)$\\
\hline
\endhead
\multicolumn{8}{c}{$N=3$}\\
\hline
0.30425     & 0.00162              & 0.30424196$^{\textrm{a)}}$ & 0.001625975$^{\textrm{a)}}$&0.2534152   & 0.0000261            & 0.2534148$^{\textrm{a)}}$  & 0.00002615$^{\textrm{a)}}$             \\
            &                      & 0.304215$^{\textrm{b)}}$   & 0.001625$^{\textrm{b)}}$   &            &                      & 0.253445$^{\textrm{b)}}$   & $<2.5\times 10^{-7}$$^{\textrm{b)}}$   \\
0.2779622   & 0.0000451            & 0.27796225$^{\textrm{a)}}$ & 0.00004505$^{\textrm{a)}}$ &0.251434    & 0.000006             & 0.251432$^{\textrm{a)}}$   & 0.00000635$^{\textrm{a)}}$             \\
            &                      & 0.27796$^{\textrm{b)}}$    & 0.000045$^{\textrm{b)}}$   &            &                      & 0.25144$^{\textrm{b)}}$    & 0.0000125$^{\textrm{b)}}$              \\
0.261020    & 0.000345             & 0.2610194$^{\textrm{a)}}$  & 0.000346$^{\textrm{a)}}$   &0.245926    & 0.000179             & 0.24615$^{\textrm{b)}}$    & 0.0006$^{\textrm{b)}}$                 \\
            &                      & 0.260975$^{\textrm{b)}}$   & 0.000325$^{\textrm{b)}}$   &0.244124    & 0.000137             & 0.24495$^{\textrm{b)}}$    & 0.000875$^{\textrm{b)}}$               \\
0.257874    & 0.000236             & 0.2578735$^{\textrm{a)}}$  & 0.0002365$^{\textrm{a)}}$  & &&&\\
\end{longtable}}}  

{\centering
{
\footnotesize
\begin{longtable}{ll|ll||ll|ll}
\caption{\label{TabF} $^3F^\textrm{o}$ resonances of helium from below the $N=2$ to below the $N=5$ threshold compared with other theoretical results.}\vspace{-23pt}\\
\hline
\multicolumn{2}{c|}{This work}&\multicolumn{2}{c||}{other results}&\multicolumn{2}{c|}{This work (continued)}&\multicolumn{2}{c}{other results (continued)}\\
$\textrm{Re}(-E)$&$\textrm{Im}(-E)$&$\textrm{Re}(-E)$&$\textrm{Im}(-E)$&$\textrm{Re}(-E)$&$\textrm{Im}(-E)$&$\textrm{Re}(-E)$&$\textrm{Im}(-E)$\\
\hline
\endfirsthead\\
\caption[]{(continued)}\vspace{-24pt}\\
\multicolumn{2}{c|}{This work}&\multicolumn{2}{c||}{other results}&\multicolumn{2}{c|}{This work (continued)}&\multicolumn{2}{c}{other results (continued)}\\
$\textrm{Re}(-E)$&$\textrm{Im}(-E)$&$\textrm{Re}(-E)$&$\textrm{Im}(-E)$&$\textrm{Re}(-E)$&$\textrm{Im}(-E)$&$\textrm{Re}(-E)$&$\textrm{Im}(-E)$\\
\hline
\endhead
\multicolumn{8}{c}{$N=3$}\\
\hline
0.3316527  & 0.0000156  & 0.33165270$^{\textrm{a)}}$& 0.000015645$^{\textrm{a)}}$            &0.2611251  & 0.0000478  & 0.2611250$^{\textrm{a)}}$ & 0.000047775$^{\textrm{a)}}$            \\
0.2709720  & 0.0000035  & 0.2709720$^{\textrm{a)}}$ & 0.0000035$^{\textrm{a)}}$              &0.2538533  & 0.0000128  & 0.25385285$^{\textrm{a)}}$& 0.0000128$^{\textrm{a)}}$              \\
0.2674098  & 0.0000025  & 0.26740965$^{\textrm{a)}}$& 0.00000254$^{\textrm{a)}}$             &&&&\\
\hline
\multicolumn{8}{c}{$N=4$}\\
\hline
0.194733    & 0.001445    & 0.1947325$^{\textrm{a)}}$ & 0.001445$^{\textrm{a)}}$  & 0.1620777   & 0.0000533   & 0.16207765$^{\textrm{a)}}$& 0.00005325$^{\textrm{a)}}$\\
            &             & 0.194725$^{\textrm{b)}}$  & 0.00145$^{\textrm{b)}}$   &             &             & 0.16208$^{\textrm{b)}}$   & 0.0000515$^{\textrm{b)}}$ \\
0.1824801   & 0.0004101   & 0.18248005$^{\textrm{a)}}$& 0.00041015$^{\textrm{a)}}$& 0.1618383   & 0.0007290   & 0.161838$^{\textrm{a)}}$  & 0.0007285$^{\textrm{a)}}$ \\
\end{longtable}}}  

{\centering
{
\footnotesize
\begin{longtable}{ll|ll||ll|ll}
\caption{\label{TabG} $^1G^\textrm{e}$ resonances of helium from below the $N=3$ to below the $N=5$ threshold compared with other theoretical results.}\vspace{-23pt}\\
\hline
\multicolumn{2}{c|}{This work}&\multicolumn{2}{c||}{other results}&\multicolumn{2}{c|}{This work (continued)}&\multicolumn{2}{c}{other results (continued)}\\
$\textrm{Re}(-E)$&$\textrm{Im}(-E)$&$\textrm{Re}(-E)$&$\textrm{Im}(-E)$&$\textrm{Re}(-E)$&$\textrm{Im}(-E)$&$\textrm{Re}(-E)$&$\textrm{Im}(-E)$\\
\hline
\endfirsthead\\
\caption[]{(continued)}\vspace{-24pt}\\
\multicolumn{2}{c|}{This work}&\multicolumn{2}{c||}{other results}&\multicolumn{2}{c|}{This work (continued)}&\multicolumn{2}{c}{other results (continued)}\\
$\textrm{Re}(-E)$&$\textrm{Im}(-E)$&$\textrm{Re}(-E)$&$\textrm{Im}(-E)$&$\textrm{Re}(-E)$&$\textrm{Im}(-E)$&$\textrm{Re}(-E)$&$\textrm{Im}(-E)$\\
\hline
\endhead
\multicolumn{8}{c}{$N=3$}\\
\hline
0.30713      & 0.00331      & 0.3071299$^{\textrm{a)}}$    & 0.003308925$^{\textrm{a)}}$    & 0.24370502   & 0.00000157   & 0.243708$^{\textrm{a)}}$     &                                \\
0.26183      & 0.00127      & 0.2618324$^{\textrm{a)}}$    & 0.001271875$^{\textrm{a)}}$    & 0.242930     & 0.000017     & 0.242965$^{\textrm{a)}}$     & 0.000015$^{\textrm{a)}}$       \\
0.255154     & 0.000025     & 0.255153555$^{\textrm{a)}}$  & 0.0000245825$^{\textrm{a)}}$   & 0.24157466   & 0.00000018   & 0.2415795$^{\textrm{a)}}$    &                                \\
0.246448     & 0.000705     & 0.246451$^{\textrm{a)}}$     & 0.00070725$^{\textrm{a)}}$     &  &&&\\
\hline
\multicolumn{8}{c}{$N=4$}\\
\hline
0.19028    & 0.00220    & 0.1902843$^{\textrm{a)}}$  & 0.002200775$^{\textrm{a)}}$ & 0.152262   & 0.000606   & 0.152258$^{\textrm{a)}}$   & 0.00060575$^{\textrm{a)}}$  \\
           &            & 0.190305$^{\textrm{b)}}$   & 0.002475$^{\textrm{b)}}$    &            &            & 0.15225$^{\textrm{b)}}$    & 0.00061$^{\textrm{b)}}$     \\
\end{longtable}}}  

{\centering
{
\footnotesize
\begin{longtable}{ll|ll||ll|ll}
\caption{\label{TabH} $^3G^\textrm{e}$ resonances of helium from below the $N=3$ to below the $N=5$ threshold compared with other theoretical results.}\vspace{-23pt}\\
\hline
\multicolumn{2}{c|}{This work}&\multicolumn{2}{c||}{other results}&\multicolumn{2}{c|}{This work (continued)}&\multicolumn{2}{c}{other results (continued)}\\
$\textrm{Re}(-E)$&$\textrm{Im}(-E)$&$\textrm{Re}(-E)$&$\textrm{Im}(-E)$&$\textrm{Re}(-E)$&$\textrm{Im}(-E)$&$\textrm{Re}(-E)$&$\textrm{Im}(-E)$\\
\hline
\endfirsthead\\
\caption[]{(continued)}\vspace{-24pt}\\
\multicolumn{2}{c|}{This work}&\multicolumn{2}{c||}{other results}&\multicolumn{2}{c|}{This work (continued)}&\multicolumn{2}{c}{other results (continued)}\\
$\textrm{Re}(-E)$&$\textrm{Im}(-E)$&$\textrm{Re}(-E)$&$\textrm{Im}(-E)$&$\textrm{Re}(-E)$&$\textrm{Im}(-E)$&$\textrm{Re}(-E)$&$\textrm{Im}(-E)$\\
\hline
\endhead
\multicolumn{8}{c}{$N=3$}\\
\hline
0.26956941   & 0.00000035   & 0.269569262$^{\textrm{a)}}$  & 0.00000036375$^{\textrm{a)}}$  & 0.24950880   & 0.00000026   & 0.249509$^{\textrm{a)}}$     &                                \\
             &              & 0.2695675$^{\textrm{b)}}$    &                                & 0.2452023    & 0.0000285    & 0.2452025$^{\textrm{a)}}$    & 0.00002625$^{\textrm{a)}}$     \\
0.259147     & 0.000054     & 0.25914687$^{\textrm{a)}}$   & 0.0000536375$^{\textrm{a)}}$   & 0.24365020   & 0.00000181   & 0.24365215$^{\textrm{a)}}$   &                                \\
             &              & 0.259145$^{\textrm{b)}}$     &                                & 0.241585014  & 0.000000155  & 0.24159$^{\textrm{a)}}$      &                                \\
\hline
\multicolumn{8}{c}{$N=4$}\\
\hline
0.181009   & 0.000121   & 0.181008965$^{\textrm{a)}}$& 0.0001211425$^{\textrm{a)}}$&0.151886   & 0.000065   & 0.151884$^{\textrm{a)}}$   & 0.00006475$^{\textrm{a)}}$  \\
           &            & 0.181007$^{\textrm{c)}}$   & 0.0001205$^{\textrm{c)}}$   &           &            & 0.151885$^{\textrm{c)}}$   & 0.0000625$^{\textrm{c)}}$   \\
\end{longtable}}}

\ack
This work was supported by the Deutsche Forschungsgemeinschaft under Contract No. FR 591/16-1 and and No. MA 3305/2-2. Access to the computing facilities of the Leibniz-Rechenzentrum der Bayerischen Akademie der Wissenschaft is gratefully acknowledged. We thank the Universit\'e catholique de Louvain for providing access to the supercomputer of the Calcul Intensif et Stockage de Masse.

\datatables
\centering
{\scriptsize
\begin{longtable}{lll|ll||lll|ll}
\caption{\label{Tab1}$^1P^\textrm{o}$ resonances of helium from below the $N=2$ to below the $N=5$ threshold compared with data from \cite{RSDK:JPB30-4663}.}\vspace{-9pt}\\
\hline
\multicolumn{3}{c|}{This work}&\multicolumn{2}{c||}{Rost {\it et al.} \cite{RSDK:JPB30-4663}}&\multicolumn{3}{c|}{This work (continued)}&\multicolumn{2}{c}{Rost {\it et al.} \cite{RSDK:JPB30-4663} (continued)}\\
$\textrm{Re}(-E)$&$\textrm{Im}(-E)$&$\langle\cos(\theta_{12})\rangle$ &$\textrm{Re}(-E)$&$\textrm{Im}(-E)$&$\textrm{Re}(-E)$&$\textrm{Im}(-E)$&$\langle\cos(\theta_{12})\rangle$&$\textrm{Re}(-E)$&$\textrm{Im}(-E)$\\
\hline
\endfirsthead\\
\caption[]{(continued)}\vspace{-11pt}\\
\multicolumn{3}{c|}{This work}&\multicolumn{2}{c||}{Rost {\it et al.} \cite{RSDK:JPB30-4663}}&\multicolumn{3}{c|}{This work (continued)}&\multicolumn{2}{c}{Rost {\it et al.} \cite{RSDK:JPB30-4663} (continued)}\\
$\textrm{Re}(-E)$&$\textrm{Im}(-E)$&$\langle\cos(\theta_{12})\rangle$ &$\textrm{Re}(-E)$&$\textrm{Im}(-E)$&$\textrm{Re}(-E)$&$\textrm{Im}(-E)$&$\langle\cos(\theta_{12})\rangle$&$\textrm{Re}(-E)$&$\textrm{Im}(-E)$\\
\hline
\endhead
\multicolumn{10}{c}{$N=2$}\\
\hline
0.69313        & 0.00069            & \wm0.03312    & 0.693134920 & 0.000686625          & 0.50641378     & 0.00000556         &  \wm0.027821  & 0.50641384  & 0.00000559\\
0.597073801    & 0.000001922        &   -0.4168372  & 0.597073804 & 0.000001923          & 0.50584264     & $<0.1\times10^{-7}$&  \wm0.25767   & 0.50584269  & \\
0.5640830      & 0.0001501          & \wm0.015690   & 0.564085188 & 0.000150594          & 0.505806932    & 0.000000047        &   -0.28819    & 0.50580696  & 0.00000010\\
0.54709214     & $<0.1\times10^{-7}$& \wm0.13595    & 0.547092709 & 0.000000005          & 0.50517456     & 0.00000404         &  \wm0.02796   & 0.505175    & 0.000004\\
0.54649297     & 0.00000096         &   -0.21408      & 0.546493257 & 0.000001014        & 0.50475851     & $<0.1\times10^{-7}$&  \wm0.26039   & 0.5047590   & \\
0.5343622      & 0.0000639          & \wm0.022382   & 0.534363144 & 0.000064173          & 0.504732103    & 0.000000035        &    -0.29033   &&\\
0.52761598     & $<0.7\times10^{-7}$& \wm0.19787    & 0.527616338 & $0.7\times10^{-10}$  & 0.50426276     & 0.00000303         &  \wm0.028056  &&\\
\hline\hline
\end{longtable}}\newpage
{\scriptsize
\begin{longtable}{lll||lll||lll}
\caption{\label{Tab2}$^3P^\textrm{o}$ resonances of helium from below the $N=2$ to below the $N=5$ threshold.}\vspace{-9pt}\\
\hline
$\textrm{Re}(-E) $&$\textrm{Im}(-E) $&$\langle\cos(\theta_{12})\rangle$&$\textrm{Re}(-E) $&$\textrm{Im}(-E) $&$\langle\cos(\theta_{12})\rangle$&$\textrm{Re}(-E) $&$\textrm{Im}(-E) $&$\langle\cos(\theta_{12})\rangle$\\
\hline
\endfirsthead\\
\caption[]{(continued)}\vspace{-11pt}\\
$\textrm{Re}(-E) $&$\textrm{Im}(-E) $&$\langle\cos(\theta_{12})\rangle$&$\textrm{Re}(-E) $&$\textrm{Im}(-E) $&$\langle\cos(\theta_{12})\rangle$&$\textrm{Re}(-E) $&$\textrm{Im}(-E) $&$\langle\cos(\theta_{12})\rangle$\\
\hline
\endhead
\multicolumn{9}{c}{$N=2$}\\
\hline
0.7604922   & 0.0001492           &    -0.364874   & 0.509123886 & 0.000001587         &    -0.42895608 & 0.50315587161& 0.00000001116      &    -0.0301797  \\ 
0.5846722   & 0.0000411           &    -0.4143051  & 0.5086964426& 0.0000000492        &    -0.02838248 & 0.5028884309 & $<0.4\times10^{-9}$&  \wm0.46023747 \\ 
0.57903096  & 0.00000094          &    -0.0208351  & 0.507508411 & $<0.2\times10^{-8}$ &  \wm0.4530691  & 0.502784604  & 0.000000267        &    -0.431226   \\ 
0.54884429  & $<0.1\times10^{-7}$ &  \wm0.3584342  & 0.507082993 & 0.000001086         &    -0.4296268  & 0.50270828176& 0.00000000890      &    -0.0303482  \\ 
0.54283722  & 0.00001576          &    -0.4215271  & 0.5067854543& 0.0000000343        &    -0.0289510  & 0.5024948854 & $<0.3\times10^{-9}$&  \wm0.46079185 \\ 
0.539558770 & 0.000000395         &    -0.023095   & 0.505960107 & $<0.1\times10^{-8}$ &  \wm0.45559221 & 0.502411338  & 0.000000216        &    -0.4313806  \\ 
0.52863838  & $<0.1\times10^{-7}$ &  \wm0.4101493  & 0.505657168  & 0.000000775        &    -0.4301258  & 0.50234959869& 0.00000000721      &    -0.0304866  \\
\hline\hline
\end{longtable}
}\newpage
{\scriptsize
\begin{longtable}{lll||lll||lll}
\caption{\label{Tab3}$^1D^\textrm{e}$ resonances of helium from below the $N=2$ to below the $N=5$ threshold.}\vspace{-9pt}\\
\hline
$\textrm{Re}(-E) $&$\textrm{Im}(-E) $&$\langle\cos(\theta_{12})\rangle$&$\textrm{Re}(-E) $&$\textrm{Im}(-E) $&$\langle\cos(\theta_{12})\rangle$&$\textrm{Re}(-E) $&$\textrm{Im}(-E) $&$\langle\cos(\theta_{12})\rangle$\\
\hline
\endfirsthead\\
\caption[]{(continued)}\vspace{-11pt}\\
$\textrm{Re}(-E) $&$\textrm{Im}(-E) $&$\langle\cos(\theta_{12})\rangle$&$\textrm{Re}(-E) $&$\textrm{Im}(-E) $&$\langle\cos(\theta_{12})\rangle$&$\textrm{Re}(-E) $&$\textrm{Im}(-E) $&$\langle\cos(\theta_{12})\rangle$\\
\hline
\endhead
\multicolumn{9}{c}{$N=2$}\\
\hline
0.70193          & 0.00118                &    -0.23441   & 0.50783793       & 0.00000084             &  \wm0.130120   & 0.502901572      & $<0.2\times 10^{-8}$   &  \wm0.263659   \\ 
0.569216         & 0.000277               &    -0.33449   & 0.507568669      & $<0.5\times10^{-8}$    &  \wm0.261873   & 0.5026694        & 0.0000024              &    -0.4029     \\ 
0.5564290        & 0.0000097              &  \wm0.0526063 & 0.50662613       & 0.00000939             &    -0.397225   & 0.50255557       & 0.00000016             &  \wm0.13859    \\ 
0.5367246        & 0.0001157              &    -0.363283  & 0.5061904        & 0.0000006              &  \wm0.132766   & 0.502505352     & $<0.1\times10^{-8}$    &  \wm0.26378    \\ 
0.5315119        & 0.0000054              &  \wm0.091932  & 0.506001475      & $<0.5\times 10^{-8}$   &  \wm0.262528   & 0.50231830       & 0.00000195             &    -0.4034     \\ 
0.52929296       & $<0.1\times10^{-7}$    &  \wm0.2485407 & 0.5053289        & 0.0000068              &    -0.39902    & 0.502225911      & 0.000000133            &  \wm0.1391     \\ 
0.5227417        & 0.0000580              &    -0.377664  & 0.50501272       & 0.00000044             &  \wm0.13465    & 0.5021850734     & $<0.7\times 10^{-9}$   &  \wm0.263882   \\
\hline\hline
\end{longtable}
}\newpage
{\scriptsize
\begin{longtable}{lll||lll||lll}
\caption{\label{Tab4}$^3D^\textrm{e}$ resonances of helium from below the $N=2$ to below the $N=5$ threshold.}\vspace{-9pt}\\
\hline
$\textrm{Re}(-E) $&$\textrm{Im}(-E) $&$\langle\cos(\theta_{12})\rangle$&$\textrm{Re}(-E) $&$\textrm{Im}(-E) $&$\langle\cos(\theta_{12})\rangle$&$\textrm{Re}(-E) $&$\textrm{Im}(-E) $&$\langle\cos(\theta_{12})\rangle$\\
\hline
\endfirsthead\\
\caption[]{(continued)}\vspace{-11pt}\\
$\textrm{Re}(-E) $&$\textrm{Im}(-E) $&$\langle\cos(\theta_{12})\rangle$&$\textrm{Re}(-E) $&$\textrm{Im}(-E) $&$\langle\cos(\theta_{12})\rangle$&$\textrm{Re}(-E) $&$\textrm{Im}(-E) $&$\langle\cos(\theta_{12})\rangle$\\
\hline
\endhead
\multicolumn{9}{c}{$N=2$}\\
\hline
0.58378424       & $<0.6\times10^{-7}$  &    -0.29371700 & 0.508085781       & 0.000000259          &  \wm0.0428806  & 0.503021708       & 0.000000061          &  \wm0.0428487  \\
0.56068624        & 0.00000371           &  \wm0.01970972 & 0.507574151       & $<0.2\times10^{-8}$  &  \wm0.3160692  & 0.5029030252      & $<0.4\times10^{-9}$  &  \wm0.3246220  \\
0.541678994       & $<0.4\times10^{-8}$  &    -0.3296968  & 0.5069578476      & $<0.4\times 10^{-9}$ &    -0.3636513  & 0.50275223380     & $<0.9\times10^{-10}$ &    -0.3685951  \\
0.53346309        & 0.00000189           &  \wm0.03657626 & 0.506364305       & 0.000000182          &  \wm0.0429305  & 0.502601511       & 0.000000048          &  \wm0.0428192  \\
0.529312233       & $<0.6\times10^{-8}$  &  \wm0.27281196 & 0.506005498       & $<0.2\times10^{-8}$  &  \wm0.3189903  & 0.5025065283      & $<0.3\times10^{-9}$  &  \wm0.3253245  \\
0.525018616       & $<0.3\times10^{-8}$  &    -0.3450135  & 0.5055664091      & $<0.3\times 10^{-9}$ &    -0.3652467  & 0.50238514134     & $<0.8\times10^{-10}$ &    -0.3690468  \\
0.521131108       & 0.000001022          &  \wm0.0408525  & 0.505139297       & 0.000000133          &  \wm0.0429297  & 0.502263235       & 0.000000039          &  \wm0.0427919  \\
\hline\hline
\end{longtable}
}
\newpage
{\scriptsize
\begin{longtable}{lll||lll||lll}
\caption{\label{Tab5}$^1F^\textrm{o}$ resonances of helium from below the $N=2$ to below the $N=5$ threshold.}\vspace{-9pt}\\
\hline
$\textrm{Re}(-E) $&$\textrm{Im}(-E) $&$\langle\cos(\theta_{12})\rangle$&$\textrm{Re}(-E) $&$\textrm{Im}(-E) $&$\langle\cos(\theta_{12})\rangle$&$\textrm{Re}(-E) $&$\textrm{Im}(-E) $&$\langle\cos(\theta_{12})\rangle$\\
\hline
\endfirsthead\\
\caption[]{(continued)}\vspace{-11pt}\\
$\textrm{Re}(-E) $&$\textrm{Im}(-E) $&$\langle\cos(\theta_{12})\rangle$&$\textrm{Re}(-E) $&$\textrm{Im}(-E) $&$\langle\cos(\theta_{12})\rangle$&$\textrm{Re}(-E) $&$\textrm{Im}(-E) $&$\langle\cos(\theta_{12})\rangle$\\
\hline
\endhead
\multicolumn{9}{c}{$N=2$}\\
\hline
0.5582817     & 0.0000061             &   -0.19531660 &0.50625337    & 0.00000038            &   -0.2243239  &0.502572180   & 0.000000104           &   -0.224432   \\ 
0.5322568     & 0.0000034             &   -0.215744   &0.506195353   & $<0.4\times 10^{-8}$  & \wm0.021908   &0.502557077   & $<0.1\times 10^{-8}$  & \wm0.021696   \\ 
0.53148168    & $<0.1\times 10^{-7}$  & \wm0.007638   &0.5060706198  & $<0.3\times 10^{-9}$  & \wm0.2006586  &0.50252384304 & $<0.8\times 10^{-10}$ & \wm0.202268   \\ 
0.5204919     & 0.0000020             &   -0.221323   &0.50505850    & 0.00000028            &   -0.2243874  &0.502239408   & 0.000000085           &   -0.2244295  \\ 
0.52012450    & $<0.1\times 10^{-7}$  & \wm0.0173120  &0.505016481   & $<0.3\times 10^{-8}$  & \wm0.021909   &0.5022271522  & $<0.8\times 10^{-9}$  & \wm0.0216449  \\ 
0.5194017125  & $<0.4\times 10^{-9}$  & \wm0.1937084  &0.5049254775  & $<0.2\times 10^{-9}$  & \wm0.2011975  &0.50220012171 & $<0.6\times 10^{-10}$ & \wm0.202405   \\ 
0.51416744    & 0.00000120            &   -0.223174   &0.504176068   & 0.000000212           &   -0.2244165  &0.501967273   & 0.000000070           &   -0.2244260  \\
\hline\hline
\end{longtable}
}\newpage
{\scriptsize
\begin{longtable}{lll||lll||lll}
\caption{\label{Tab6}$^3F^\textrm{o}$ resonances of helium from below the $N=2$ to below the $N=5$ threshold.}\vspace{-9pt}\\
\hline
$\textrm{Re}(-E) $&$\textrm{Im}(-E) $&$\langle\cos(\theta_{12})\rangle$&$\textrm{Re}(-E) $&$\textrm{Im}(-E) $&$\langle\cos(\theta_{12})\rangle$&$\textrm{Re}(-E) $&$\textrm{Im}(-E) $&$\langle\cos(\theta_{12})\rangle$\\
\hline
\endfirsthead\\
\caption[]{(continued)}\vspace{-11pt}\\
$\textrm{Re}(-E) $&$\textrm{Im}(-E) $&$\langle\cos(\theta_{12})\rangle$&$\textrm{Re}(-E) $&$\textrm{Im}(-E) $&$\langle\cos(\theta_{12})\rangle$&$\textrm{Re}(-E) $&$\textrm{Im}(-E) $&$\langle\cos(\theta_{12})\rangle$\\
\hline
\endhead
\multicolumn{9}{c}{$N=2$}\\
\hline
0.56620295    & 0.00001166             &   -0.2585941 & 0.506561288   & 0.000000447            &   -0.2897877 & 0.502652743   & 0.000000116          &   -0.2963643 \\
0.5358282     & 0.0000053              &   -0.2602169 & 0.506196653   & 0.000000007            & \wm0.0841986 & 0.5025574510  & 0.0000000021         & \wm0.0909389 \\
0.531488382   & 0.000000032            & \wm0.0311738 & 0.5060706454  & $<0.1\times 10^{-9}$   & \wm0.2025720 & 0.50252385132 & $<0.4\times 10^{-10}$& \wm0.2046521 \\
0.52232617    & 0.00000271             &   -0.2701480 & 0.505282069   & 0.000000323            &   -0.291842  & 0.5023047778  & 0.0000000938         &   -0.296998  \\
0.520129747   & 0.000000027            & \wm0.0562875 & 0.505017454   & 0.000000005            & \wm0.0864094 & 0.5022274580  & 0.0000000017         & \wm0.091537  \\
0.5194017624  & $<0.2\times 10^{-9}$   & \wm0.1943005 & 0.50492549744 & $<0.8\times 10^{-10}$  & \wm0.2032566 & 0.50220012855 & $<0.3\times 10^{-10}$& \wm0.2048345 \\
0.51522352    & 0.00000155             &   -0.277824  & 0.504343460   & 0.000000242            &   -0.2934067 & 0.5020210412  & 0.0000000770         &   -0.297523  \\
\hline\hline
\end{longtable}
}
\newpage
{\scriptsize
\begin{longtable}{lll||lll||lll}
\caption{\label{Tab7}$^1G^\textrm{e}$ resonances of helium from below the $N=2$ to below the $N=5$ threshold.}\vspace{-9pt}\\
\hline
$\textrm{Re}(-E) $&$\textrm{Im}(-E) $&$\langle\cos(\theta_{12})\rangle$&$\textrm{Re}(-E) $&$\textrm{Im}(-E) $&$\langle\cos(\theta_{12})\rangle$&$\textrm{Re}(-E) $&$\textrm{Im}(-E) $&$\langle\cos(\theta_{12})\rangle$\\
\hline
\endfirsthead\\
\caption[]{(continued)}\vspace{-11pt}\\
$\textrm{Re}(-E) $&$\textrm{Im}(-E) $&$\langle\cos(\theta_{12})\rangle$&$\textrm{Re}(-E) $&$\textrm{Im}(-E) $&$\langle\cos(\theta_{12})\rangle$&$\textrm{Re}(-E) $&$\textrm{Im}(-E) $&$\langle\cos(\theta_{12})\rangle$\\
\hline
\endhead
\multicolumn{9}{c}{$N=2$}\\
\hline
0.53235728     & 0.00000005     &   -0.16739070 &  0.50617945015  & 0.00000000005  & \wm0.0435745  & 0.50293654488  & 0.00000000001  & \wm0.1533523  \\
0.52057826     & 0.00000004     &   -0.1804131  &  0.50610643795  & 0.00000000003  & \wm0.1536049  & 0.5025780422   & 0.0000000033   &   -0.2054725  \\
0.5200335936   & 0.0000000001   & \wm0.0118889  &  0.505073969    & 0.000000009    &   -0.2014504  & 0.50255283555  & 0.00000000001  & \wm0.051516   \\
0.514226954    & 0.000000031    &   -0.1883773  &  0.50500487157  & 0.00000000004  & \wm0.0461648  & 0.50253337631  & 0.00000000001  & \wm0.1533140  \\
0.51390959967  & 0.00000000010  & \wm0.0261414  &  0.50495159637  & 0.00000000002  & \wm0.1535207  & 0.5022441968   & 0.0000000027   &   -0.2060317  \\
0.51366442249  & 0.00000000004  & \wm0.15406527 &  0.504187859    & 0.000000007    &   -0.2028490  & 0.50222370233  & 0.00000000001  & \wm0.052228   \\
0.510418204    & 0.000000022    &   -0.1935520  &  0.50413592059  & 0.00000000003  & \wm0.0480702  & 0.50220787551  & 0.00000000001  & \wm0.1532816  \\
\hline\hline
\end{longtable}
}\newpage
{\scriptsize
\begin{longtable}{lll||lll||lll}
\caption{\label{Tab8}$^3G^\textrm{e}$ resonances of helium from below the $N=2$ to below the $N=5$ threshold.}\vspace{-9pt}\\
\hline
$\textrm{Re}(-E) $&$\textrm{Im}(-E) $&$\langle\cos(\theta_{12})\rangle$&$\textrm{Re}(-E) $&$\textrm{Im}(-E) $&$\langle\cos(\theta_{12})\rangle$&$\textrm{Re}(-E) $&$\textrm{Im}(-E) $&$\langle\cos(\theta_{12})\rangle$\\
\hline
\endfirsthead\\
\caption[]{(continued)}\vspace{-11pt}\\
$\textrm{Re}(-E) $&$\textrm{Im}(-E) $&$\langle\cos(\theta_{12})\rangle$&$\textrm{Re}(-E) $&$\textrm{Im}(-E) $&$\langle\cos(\theta_{12})\rangle$&$\textrm{Re}(-E) $&$\textrm{Im}(-E) $&$\langle\cos(\theta_{12})\rangle$\\
\hline
\endhead
\multicolumn{9}{c}{$N=2$}\\
\hline
0.532368759   & 0.000000054    &   -0.16608372 & 0.50617945107 & 0.00000000003  & \wm0.0436364  & 0.50293654487 & 0.00000000001  & \wm0.1533507  \\
0.520586136   & 0.000000045    &   -0.1794857  & 0.50610643797 & 0.00000000003  & \wm0.1536047  & 0.5025784691  & 0.0000000034   &   -0.2054258  \\
0.5200336004  & 0.0000000001   & \wm0.0119605  & 0.5050751330  & 0.000000009    &   -0.2013233  & 0.50255283561 & 0.00000000001  & \wm0.0515389  \\
0.514231962   & 0.000000032    &   -0.1877861  & 0.50500487210 & 0.00000000002  & \wm0.0462162  & 0.50253337631 & 0.00000000001  & \wm0.1533121  \\
0.5139096045  & 0.0000000001   & \wm0.0262304  & 0.50495159636 & 0.00000000002  & \wm0.1535201  & 0.5022445439  & 0.0000000028   &   -0.2059927  \\
0.51366442250 & 0.00000000004  & \wm0.15406575 & 0.504188737   & 0.000000007    &   -0.2027546  & 0.50222370236 & 0.00000000001  & \wm0.0522463  \\
0.510421483   & 0.000000022    &   -0.1931701  & 0.50413592090 & 0.00000000002  & \wm0.0481126  & 0.50220787550 & 0.00000000001  & \wm0.1532795  \\
\hline\hline
\end{longtable}
}\newpage
\normalsize
\bibliographystyle{prsty}
\bibliography{./bibliography}

\begin{thebibliography}{10}

\bibitem{RSDK:JPB30-4663}
J.~M. Rost, K. Schulz, M. Domke, and G. Kaindl, J. Phys. B {\bf 30},  4663
  (1997).

\bibitem{MC:PRL10-516}
R.~P. Madden and K. Codling, Phys. Rev. Lett. {\bf 10},  516  (1963).

\bibitem{RW:PRL65-1965}
K. Richter and D. Wintgen, Phys. Rev. Lett. {\bf 65},  1965  (1990).

\bibitem{RBWS:JPB25-3929}
K. Richter, J.~S. Briggs, D. Wintgen, and E.~A. Solov'ev, J. Phys. B {\bf 25},
  3929  (1992).

\bibitem{PS:Diss99}
P. Schlagheck, Dissertation, Technische Universit\"at M\"unchen, 1999.

\bibitem{SB:EPJD22-401}
P. Schlagheck and A. Buchleitner, Eur. Phys. J. D {\bf 22},  401  (2003).

\bibitem{JM:Diss04}
G.~J. {Madro$\tilde{\textrm{n}}$ero~Pab\'on}, Dissertation,
  Ludwig-Maximilians-Universit\"at M\"unchen, 2004.

\bibitem{MB:PRA77-053402}
J. Madro{\~{n}}ero and A. Buchleitner, Phys. Rev. A {\bf 77},  053402  (2008).

\bibitem{BDZ:PRep368-409}
A. Buchleitner, D. Delande, and J. Zakrzewski, Phys. Rep. {\bf 368},  409
  (2002).

\bibitem{FBCK:PRL69-2642}
D.~N. Fittinghoff, P.~R. Bolton, B. Chang, and K.~C. Kulander, Phys. Rev. Lett.
  {\bf 69},  2642  (1992).

\bibitem{WSDASK:PRL73-1227}
B. Walker {\it et~al.}, Phys. Rev. Lett. {\bf 73},  1227  (1994).

\bibitem{SYDK:PRL70-1599}
K.~J. Schafer, B. Yang, L.~F. DiMauro, and K.~C. Kulander, Phys. Rev. Lett.
  {\bf 70},  1599  (1993).

\bibitem{YSWKAD:PRL71-3770}
B. Yang {\it et~al.}, Phys. Rev. Lett. {\bf 71},  3770  (1993).

\bibitem{W+:Nat405-658}
T. Weber {\it et~al.}, Nature {\bf 405},  658  (2000).

\bibitem{M+:PRA65-035401}
R. Moshammer {\it et~al.}, Phys. Rev. A {\bf 65},  035401  (2002).

\bibitem{dJ+:JPB37-L161}
V.~L.~B. de~Jesus {\it et~al.}, J. Phys. B {\bf 37},  L161  (2004).

\bibitem{PDTSBD:PRL96-133001}
J.~S. Parker {\it et~al.}, Phys. Rev. Lett. {\bf 96},  133001  (2006).

\bibitem{NGSES:PRL101-233001}
T. Nubbemeyer {\it et~al.}, Phys. Rev. Lett. {\bf 101},  233001  (2008).

\bibitem{PMMT:JPB40-1729}
J.~S. Parker, K.~J. Meharg, G.~A. McKenna, and K.~T. Taylor, J. Phys. B {\bf
  40},  1729  (2007).

\bibitem{NL:JPB34-545}
L.~A.~A. Nikolopoulos and P. Lambropoulos, J. Phys. B {\bf 34},  545  (2001).

\bibitem{FvdH:PRA66-031402R}
L. Feng and H.~W. van~der Hart, Phys. Rev. A {\bf 66},  031402(R)  (2002).

\bibitem{FvdH:JPB36-L1}
L. Feng and H.~W. van~der Hart, J. Phys. B {\bf 36},  L1  (2003).

\bibitem{PBLB:EPJD26-7}
B. Piraux, J. Bauer, S. Laulan, and H. Bachau, Eur. Phys. J. D {\bf 26},  7
  (2003).

\bibitem{HCC:JPB38-L35}
S.~X. Hu, J. Colgan, and L.~A. Collins, J. Phys. B {\bf 38},  L35  (2005).

\bibitem{IK:PRA71-043405}
I.~A. Ivanov and A.~S. Kheifets, Phys. Rev. A {\bf 71},  043405  (2005).

\bibitem{KI:JPB39-1731}
A.~S. Kheifets and I.~A. Ivanov, J. Phys. B {\bf 39},  1731  (2006).

\bibitem{FLKEP:PRA74-063409}
E. Foumouo, G. {Lagmago Kamta}, G. Edah, and B. Piraux, Phys. Rev. A {\bf 74},
  063409  (2006).

\bibitem{F+:JPB41-051001}
E. Foumouo {\it et~al.}, J. Phys. B {\bf 41},  051001  (2008).

\bibitem{P:PRev112-1649}
C.~L. Pekeris, Phys. Rev. {\bf 112},  1649  (1958).

\bibitem{P:PRev115-1216}
C.~L. Pekeris, Phys. Rev. {\bf 115},  1216  (1959).

\bibitem{BWR:JPB28-3163}
A. B{\"u}rgers, D. Wintgen, and J.-M. Rost, J. Phys. B {\bf 28},  3163  (1995).

\bibitem{GD:EPL40-363}
B. Gr\'{e}maud and D. Delande, Europhys. Lett. {\bf 40},  363  (1997).

\bibitem{BG:Diss97}
B. Gr{\'e}maud, Th\`ese de doctorat, Universit{\'e} Pierre et Marie Curie
  (Paris 6), 1997.

\bibitem{JPDMK:PRA78-021401R}
Y.~H. Jiang {\it et~al.}, Phys. Rev. A {\bf 78},  021401(R)  (2008).

\bibitem{K:CPAM10-151}
T. Kato, Comm. Pure Appl. Math. {\bf 10},  151  (1957).

\bibitem{W:PRev122-1826}
A.~W. Weiss, Phys. Rev. {\bf 122},  1826  (1961).

\bibitem{RW:RMP32-194}
C.~C.~J. Roothaan and A.~W. Weiss, Rev. Mod. Phys. {\bf 32},  194  (1960).

\bibitem{LKPS:PRA63-040502}
G. {Lagmago Kamta}, B. Piraux, and A. Scrinzi, Phys. Rev. A {\bf 63},
  040502(R)  (2001).

\bibitem{MHRM:PRA69-032707}
C.~W. McCurdy, D.~A. Horner, T.~N. Rescigno, and F. Mart\'in, Phys. Rev. A {\bf
  69},  032707  (2004).

\bibitem{LB:PRA69-033408}
S. Laulan and H. Bachau, Phys. Rev. A {\bf 69},  033408  (2004).

\bibitem{AFPSHNM:PRA78-023415}
P. Antoine {\it et~al.}, Phys. Rev. A {\bf 78},  023415  (2008).

\bibitem{EPM:PRA80-022511}
J. Eiglsperger, B. Piraux, and J. Madro{\~n}ero, Phys. Rev. A {\bf 80},  022511
   (2009).

\bibitem{EPM:PRA81-042527}
J. Eiglsperger, B. Piraux, and J. Madro\~nero, Phys. Rev. A {\bf 81},  042527
  (2010).

\bibitem{EPM:PRA81-042528}
J. Eiglsperger, B. Piraux, and J. Madro\~nero, Phys. Rev. A {\bf 81},  042528
  (2010).

\bibitem{GLK:Diss99}
G. {Lagmago Kamta}, Ph.D. thesis, Universit{\'e} Nationale du B{\'e}nin, 1999.

\bibitem{EF:Diss08}
E. Foumouo, Dissertation doctorale, Universit{\'e} catholique de Louvain, 2008.

\bibitem{JE:Diss10}
J. Eiglsperger, Dissertation, Technische Universit\"at M\"unchen, 2010.

\bibitem{R:AAMOP6-233}
M. Rotenberg, Adv. At. Mol. Phys. {\bf 6},  233  (1970).

\bibitem{HPBG:PRA55-2132}
E. Huens, B. Piraux, A. Bugacov, and M. Gajda, Phys. Rev. A {\bf 55},  2132
  (1997).

\bibitem{VMK:QuantTheoAngMom08}
D.~A. Varschalovich, A.~N. Moskalev, and V.~K. Khersonskii, {\em Quantum Theory
  of Angular Momentum} (World Scientific, Singapore, 2008).

\bibitem{CFP:PRL10-518}
J.~W. Cooper, U. Fano, and F. Prats, Phys. Rev. Lett. {\bf 10},  518  (1963).

\bibitem{AC:CMP22-269}
J. Aguilar and J.~M. Combes, Comm. Math. Phys. {\bf 22},  269  (1971).

\bibitem{BC:CMP22-280}
E. Balslev and J.~M. Combes, Comm. Math. Phys. {\bf 22},  280  (1971).

\bibitem{S:AoM97-247}
B. Simon, Ann. Math. {\bf 97},  247  (1973).

\bibitem{R:ARPC33-223}
W.~P. Reinhardt, Annu. Rev. Phys. Chem. {\bf 33},  223  (1982).

\bibitem{H:PRep99-1}
Y.~K. Ho, Phys. Rep. {\bf 99},  1  (1983).

\bibitem{GGS:AIHP42-215}
S. Graffi, V. Grecchi, and H.~J. Silverstone, Ann. Inst. Henri Poincar{\'e},
  Sect. A {\bf 42},  215  (1985).

\bibitem{PS:PRA43-3764}
M. Pont and R. Shakeshaft, Phys. Rev. A {\bf 43},  3764  (1991).

\bibitem{SR:AnalysisOfOperators78}
M. Reed and B. Simon, {\em Methods of modern mathematical physics} (Academic
  Press, New York, 1978), Vol.~IV. Analysis of operators.

\bibitem{BGD:JPB27-2663}
A. Buchleitner, B. Gr\'emaud, and D. Delande, J. Phys. B {\bf 27},  2663
  (1994).

\bibitem{JR:PRA28-1930}
B.~R. Johnson and W.~P. Reinhardt, Phys. Rev. A {\bf 28},  1930  (1983).

\bibitem{BT:PRA11-2018}
A.~K. Bhatia and A. Temkin, Phys. Rev. A {\bf 11},  2018  (1975).

\bibitem{H:JPB12-387}
Y.~K. Ho, J. Phys. B {\bf 12},  387  (1979).

\bibitem{H:PLA79-44}
Y.~K. Ho, Phys. Lett. A {\bf 79},  44   (1980).

\bibitem{H:PRA23-2137}
Y.~K. Ho, Phys. Rev. A {\bf 23},  2137  (1981).

\bibitem{H:JPB15-L691}
Y.~K. Ho, J. Phys. B {\bf 15},  L691  (1982).

\bibitem{BBFT:JPB15-L603}
K.~A. Berrington, P.~G. Burke, W.~C. Fon, and K.~T. Taylor, J. Phys. B {\bf
  15},  L603  (1982).

\bibitem{BT:PRA29-1895}
A.~K. Bhatia and A. Temkin, Phys. Rev. A {\bf 29},  1895  (1984).

\bibitem{HC:JPB18-3481}
Y.~K. Ho and J. Callaway, J. Phys. B {\bf 18},  3481  (1985).

\bibitem{O:PRA33-824}
D.~H. Oza, Phys. Rev. A {\bf 33},  824  (1986).

\bibitem{H:ZfPD11-277}
Y.~K. Ho, Z. Phys. D {\bf 11},  277  (1989).

\bibitem{SCK:PRA39-5111}
S. Salomonson, S.~L. Carter, and H.~P. Kelly, Phys. Rev. A {\bf 39},  5111
  (1989).

\bibitem{HH:JPB22-3397}
P. Hamacher and J. Hinze, J. Phys. B {\bf 22},  3397  (1989).

\bibitem{GB:JPB23-365}
R. Gersbacher and J.~T. Broad, J. Phys. B {\bf 23},  365  (1990).

\bibitem{FI:JPB23-679}
C.~F. Fischer and M. Idrees, J. Phys. B {\bf 23},  679  (1990).

\bibitem{WX:JPB23-727}
L. Wu and J. Xi, J. Phys. B {\bf 23},  727  (1990).

\bibitem{SS:PRA42-2562}
J.~M. Seminario and F.~C. Sanders, Phys. Rev. A {\bf 42},  2562  (1990).

\bibitem{H:ZfPD21-191}
Y.~K. Ho, Z. Phys. D {\bf 21},  191  (1991).

\bibitem{BMRY:ADNDT48-167}
H. Bachau, F. Mart{\'i}n, A. Riera, and M. Y{\'a\~n}ez, At. Data Nucl. Data
  Tables {\bf 48},  167   (1991).

\bibitem{CT:PRA44-232}
T.~N. Chang and X. Tang, Phys. Rev. A {\bf 44},  232  (1991).

\bibitem{MS:PRA44-7206}
L.~W. Manning and F.~C. Sanders, Phys. Rev. A {\bf 44},  7206  (1991).

\bibitem{TWM:PRA46-2437}
J.-Z. Tang, S. Watanabe, and M. Matsuzawa, Phys. Rev. A {\bf 46},  2437
  (1992).

\bibitem{C:PRA47-705}
T.~N. Chang, Phys. Rev. A {\bf 47},  705  (1993).

\bibitem{L:PRA49-4473}
E. Lindroth, Phys. Rev. A {\bf 49},  4473  (1994).

\bibitem{TS:PRA50-1321}
J.-Z. Tang and I. Shimamura, Phys. Rev. A {\bf 50},  1321  (1994).

\bibitem{DSRKW:PRA53-1424}
M. Domke {\it et~al.}, Phys. Rev. A {\bf 53},  1424  (1996).

\bibitem{H:ZfPD42-77}
Y.~K. Ho, Z. Phys. D {\bf 42},  77  (1997).

\bibitem{C:PRA56-4537}
M.-K. Chen, Phys. Rev. A {\bf 56},  4537  (1997).

\bibitem{AM:JPB40-3655}
L. Argenti and R. Moccia, J. Phys. B {\bf 40},  3655  (2007).

\bibitem{H:JPB17-L695}
Y.~K. Ho, J. Phys. B {\bf 17},  L695  (1984).

\bibitem{SM:PRA47-1878}
I. S\'anchez and F. Martin, Phys. Rev. A {\bf 47},  1878  (1993).

\bibitem{H:PRA48-3598}
Y.~K. Ho, Phys. Rev. A {\bf 48},  3598  (1993).

\bibitem{A:ADNDT94-903}
L. Argenti, At. Data Nucl. Data Tables {\bf 94},  903   (2008).

\bibitem{HB:PRA44-2895}
Y.~K. Ho and A.~K. Bhatia, Phys. Rev. A {\bf 44},  2895  (1991).

\bibitem{BH:JPB31-3307}
A.~K. Bhatia and Y.~K. Ho, J. Phys. B {\bf 31},  3307  (1998).

\bibitem{IH:CJP39-415}
I. Ivanov and Y.~K. Ho, Chin. J. Phys. {\bf 39},  415  (2001).

\end{thebibliography}

\end{document}